\documentclass[aps,pra,twocolumn,superscriptaddress,showpacs]{revtex4}

\usepackage{graphicx}
\usepackage{natbib}
\usepackage{bbold}
\newcommand{\ket}[1]{|{#1}\rangle}

\newcommand{\braket}[1]{\langle{#1} \rangle}

\newcommand{\lev}[3]{\textit{#1}$_{#2/#3}$}

\begin{document}

\title{Transmission spectroscopy of a single atom in the presence of tensor light shifts}
\author{Matthias Steiner}
\affiliation{Centre for Quantum Technologies, 3 Science Drive 2, Singapore 117543}
\affiliation{Department of Physics, National University of Singapore, 2 Science Drive 3, Singapore 117542}
\author{Yue-Sum Chin}
\affiliation{Centre for Quantum Technologies, 3 Science Drive 2, Singapore 117543}
\author{Christian Kurtsiefer}
\email{christian.kurtsiefer@gmail.com}
\affiliation{Centre for Quantum Technologies, 3 Science Drive 2, Singapore 117543}
\affiliation{Department of Physics, National University of Singapore, 2 Science Drive 3, Singapore 117542}
\date{\today}

\begin{abstract}
We investigate the interplay between Zeeman and light shifts in the transmission spectrum of an optically trapped, spin-polarized Rubidium atom. 
The spectral shape of the transmission changes from  multiple, broad resonances to a single, narrow Lorentzian with a high resonant extinction value when we increase the magnetic field strength and lower the depth of the dipole trap.  
We present an experimental configuration well-suited for quantum information applications in that it enables not only efficient light-atom coupling but also a long coherence time between ground state hyperfine levels.
\end{abstract}


\maketitle
\section{Introduction}
Individually controlled neutral atoms have been established as a viable platform for advanced applications in quantum information science~\cite{Bernien2017,Lienhard2018}.  
In this approach, a qubit is typically realized by two ground state hyperfine levels of the atom.  
Several strategies have been developed to connect multiple atomic qubits. 
For example, nearby atoms can interact by optical coupling to highly-excited Rydberg states~\cite{Wilk2010}.
Alternatively, atoms separated by large distances can be connected through an
optical link and the exchange of single photons~\cite{Ritter2012}. 
In both cases, efficient and well-controlled coupling of optical fields to the atoms is essential for using neutral atoms for quantum information applications.    
Depending on the experimental configuration, however, the conditions necessary
for efficient optical coupling can compromise other qubit properties such as the coherence time. 

We investigate this trade-off for an approach where efficient light-atom coupling is achieved by trapping individual atoms in optical tweezers and placing them at the focal spot of a high numerical aperture lens. 
In previous work, we used such an arrangement to realize strong extinction of a coherent beam by a single atom~\cite{Tey2008,Chin2017a} and resolve scattering dynamics for various temporal profiles of the incident light~\cite{Aljunid2013,Leong2016,Steiner2017}. 
In this work, we show that conditions for efficient light-atom interaction, i.e. strong extinction, are compatible with a long qubit coherence time.
In contrast to the experiments in
\cite{Tey2008,Chin2017a,Aljunid2013,Leong2016,Steiner2017}, here we use a
linearly polarized dipole trap, which strongly reduces atomic motion induced qubit dephasing but affects the light-atom coupling through a tensor light shift. 
We perform transmission spectroscopy to investigate the impact of the tensor light shift on the optical coupling in detail.
\begin{figure} 
\centering
  \includegraphics[width=\columnwidth]{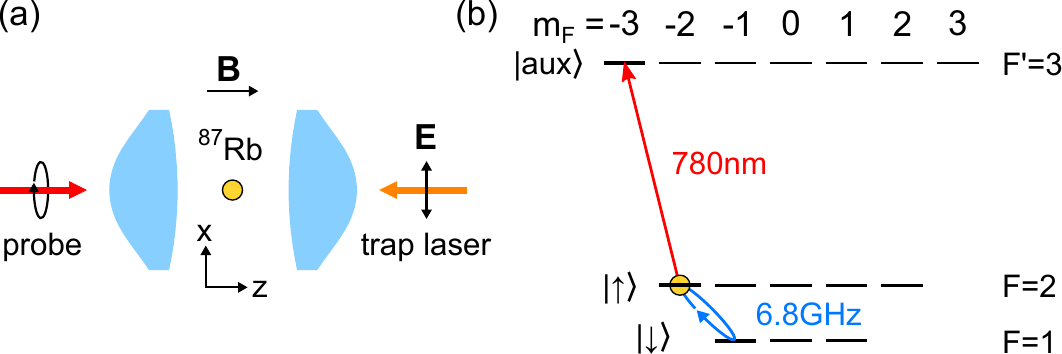}
  \caption{\label{fig:geometry} (a) Simplified optical setup. A single atom is held by a linearly polarized FORT and is probed with a circularly polarized beam. 
  (b) Energy level scheme. The two 5\lev{S}{1}{2} Zeeman levels $\ket{F=2,m=-2}\equiv\ket{\uparrow}$ and $\ket{F=1,m=-1}\equiv\ket{\downarrow}$ are used as qubit states.  
  The $\ket{\uparrow}$ state is coupled via a closed optical transition to the 5\lev{P}{3}{2} $\ket{F'=3,m=-3}\equiv\ket{\textrm{aux}}$ state. 
}
\end{figure}

\section{Zeeman and light shift Hamiltonian}
We consider an optically trapped $^{87}$Rb atom in a magnetic field applied along the quantization axis [z-axis, see Fig.~\ref{fig:geometry}(a)]. 
The magnetic field lifts the degeneracy of the Zeeman levels with the corresponding Hamiltonian~$\hat{H}_\textrm{B,F}$ for the hyperfine manifold~$F$,
\begin{equation}\label{eq:zeeman}
\hat{H}_\textrm{B,F} =  \hbar \omega_{L} \hat{F}_{z} \,,
\end{equation}
where $ \omega_{L}$  and $\hat{F}_{z}$ are the Larmor frequency and the z-component of the total angular momentum operator~$\hat{\textbf{F}}$  of the respective hyperfine level.
We use the two 5\lev{S}{1}{2} Zeeman levels  $\ket{F=2,m=-2}\equiv\ket{\uparrow}$ and $\ket{F=1,m=-1}\equiv\ket{\downarrow}$  as qubit states~(Fig.~\ref{fig:geometry}b). 
The choice of these states over the commonly used clock states $\ket{F=2,m=0}$ and $\ket{F=1,m=0}$ is motivated by the possibility to couple $\ket{\uparrow}$ via a closed optical transition to the 5\lev{P}{3}{2} $\ket{F'=3,m=-3}\equiv\ket{\textrm{aux}}$ state. 

The energy levels are further shifted by the light shift induced by the trapping field. 
For each hyperfine manifold, the light-shift Hamiltonian~$\hat{H}_\textrm{ls,F}$ can be decomposed into a scalar, a vector, and a tensor term,
\begin{equation}\label{eq:Hls}
 \hat{H}_\textrm{ls,F}= U_0 \left( c_{s} + c_{v} \left(\epsilon^* \times \epsilon\right) \hat{\textbf{F}} + c_{t} \left|\epsilon \cdot \hat{\textbf{F}}\right|^2  \right)\,,
\end{equation}
where $U_0$ is the trap depth, $\epsilon$ is the polarization vector of the trapping field, and $c_{s}$, $c_{v}$, and $c_{t}$ are the coefficients of the scalar, vector, and tensor light shifts~\cite{Deutsch2010}. 

The qubit coherence is greatly affected when the trapping field causes a frequency shift~$\delta$ of the $\ket{\uparrow}$ to $\ket{\downarrow}$  transition. 
Then the qubit frequency changes as the atom oscillates in the trap, which leads to dephasing on a timescale of $T^*_2\approx \frac{U_0}{\delta \pi k_B T_\textrm{atom}}$ ($T_\textrm{atom}$ is the temperature of the atom and $k_B$ is the Boltzmann constant)~\cite{Romalis1999}.
In a far off-resonant dipole trap~(FORT), the contribution of the scalar and the tensor term to the shift~$\delta$ is negligible~($c_{s}$ is the same for the two ground state hyperfine manifolds  5\lev{S}{1}{2}  $F=1,2$ and $c_{t}\approx  0$).  
The vector light shift, however, leads to rapid dephasing.
For example, a 1\,mK-deep, circularly polarized trap at 851\,nm shifts the qubit frequency by $\delta=2.6$\,MHz;
thus for a typical atom temperature of $50\,\mu$K, the dephasing time $T^*_2\approx2\,\mu$s is prohibitively short for quantum information purposes. 
Therefore, we use a FORT linearly polarized along the x-axis, for which the vector shift vanishes~$\left(\epsilon^* \times \epsilon=0\right)$. 
In this configuration the light-shift Hamiltonian for the excited state hyperfine manifold 5\lev{P}{3}{2}~$F'=3$ reads
\begin{equation}\label{eq:Hex}
 \hat{H}_\textrm{ls,F'=3}= U_0 \left( c_{s}  + c_{t} F^2_x  \right)\,,
\end{equation}
with $c_{s}= 0.7417$ and $c_{t}= -0.0716$ for a FORT operating at 851\,nm.
The nonlinear term proportional to $F^2_x$ leads to energy eigenstates that are
superpositions of either even or odd $m_z$ states. 
Consequently, the absorption spectrum and, in particular, the optical coupling
between~$\ket{\uparrow}$ and $\ket{\textrm{aux}}$ depend strongly on the relative strength of Zeeman and light shift. 

\section{Transmission experiment}
\begin{figure} 
\centering
  \includegraphics[width=\columnwidth]{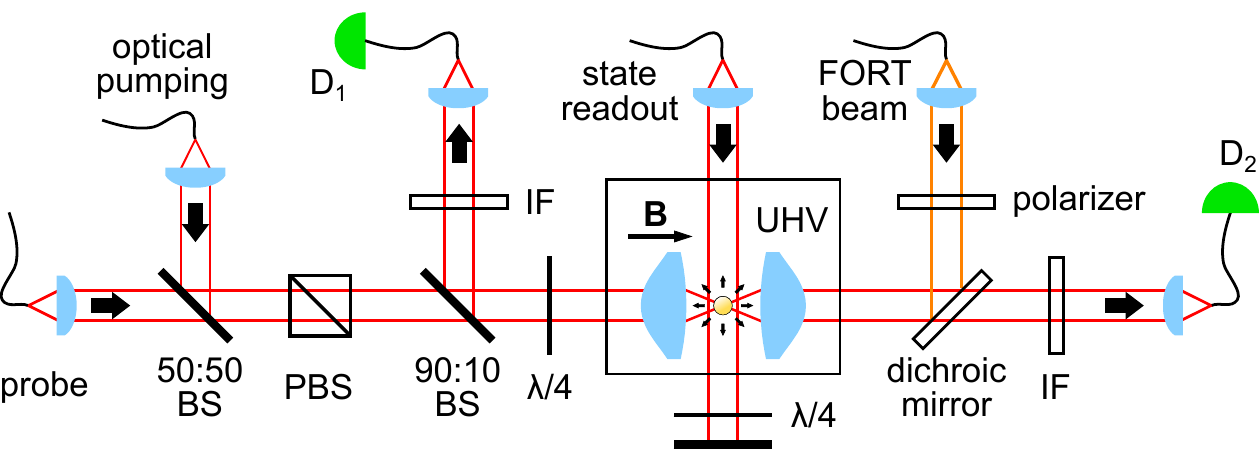}
  \caption{\label{fig:setup}  Optical setup. 
  D$_{1(2)}$:~single mode fiber connected to avalanche photodetector, (P)BS: (polarizing) beam splitter, $\lambda/4$:~quarter-wave plate, IF: interference filter. 
}
\end{figure}

To determine the impact of the light shift on the optical coupling, we perform transmission spectroscopy on a single $^{87}$Rb atom in a tightly focused red-detuned FORT~\cite{Schlosser2001}. 
The atom is held between two high numerical aperture lenses (NA=0.75, focal length~$f$=5.95\,mm) with a 2.24\,mK-deep FORT operating at a wavelength 851\,nm~\cite{Chin2017,Chin2017a}. 
Part of the atomic fluorescence is collected by the same lenses and coupled to single mode fibers connected to avalanche photodetectors, $D_1$ and $D_2$~(Fig.~\ref{fig:setup}). 

After loading an atom into the FORT, we cool the atom to 16.4(6)\,$\mu$K by 10\,ms of polarization gradient cooling~\cite{Chin2017b}. 
Then, a bias magnetic field is applied along the quantization axis (z-axis), and the atom is optically pumped into $\ket{\uparrow}$. 
We probe the light-atom interaction with a circularly polarized ($\sigma^-$) beam, driving the transition $\ket{\uparrow}$ to $\ket{\textrm{aux}}$ near 780\,nm. 
The Rabi frequency of the driving field $\Omega = 0.052(3)\Gamma$ is set far below saturation~($\Gamma=2\pi\times 6.07\,$MHz is the spontaneous decay rate). 
During the 1\,ms-long probe pulse, we accumulate the number of detected photons~$n_\textrm{p}$ at the detector~$D_2$.  
We then obtain the transmission $T=n_\textrm{p}/n_0$ by comparing $n_\textrm{p}$ to the number of detected photons in a reference measurement~$n_0$ during which the atom is in a state off-resonant with the probe field (F=1). 
\begin{figure} 
\centering
  \includegraphics[width=\columnwidth]{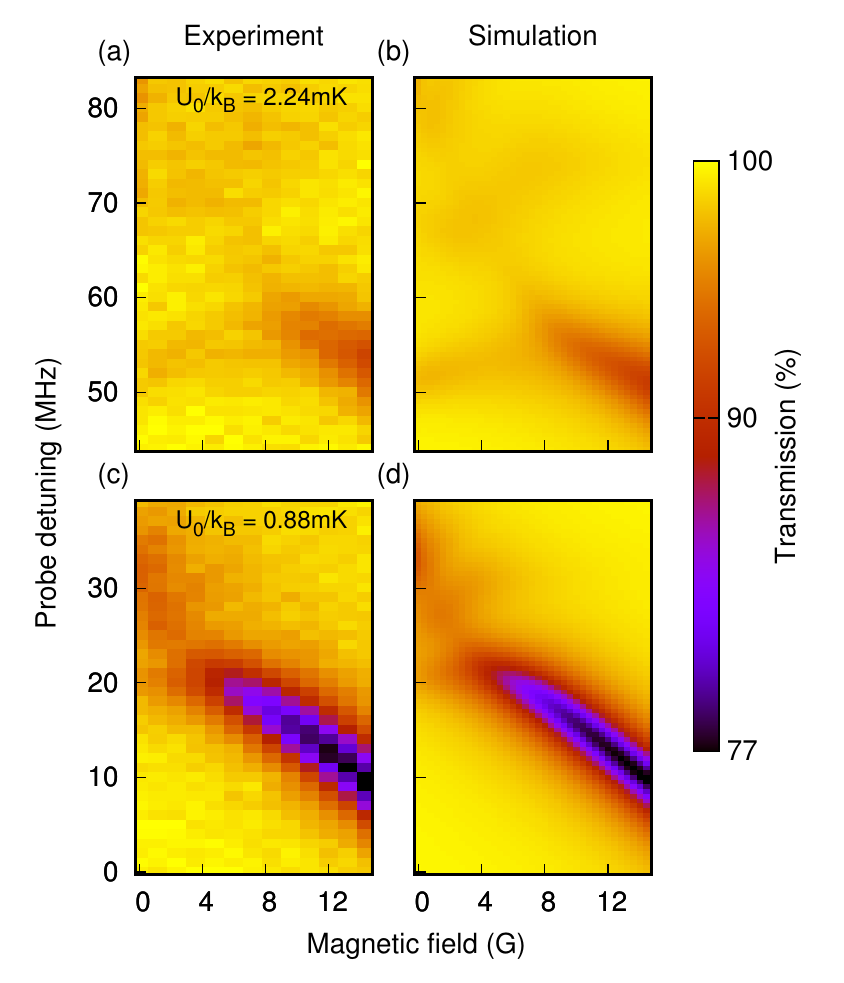}
  \caption{\label{fig:tx_matrix} Transmission spectra of a single atom in a deep (a,b) and a shallow (c,d) trap. 
  For weak magnetic fields, the light shift of the FORT leads to strong state mixing in the excited state.
  In the shallow trap and with a strong magnetic field applied, the probe field
  couples efficiently to the transition between $\ket{\uparrow}$ and
  $\ket{\textrm{aux}}$ and a high extinction ($\approx23\%$) is observed. }
\end{figure}

Figure~\ref{fig:tx_matrix}(a) shows the observed transmission spectrum as we vary the frequency of the probe field and the amplitude of the bias magnetic field. 
We observe a peak extinction of $\epsilon= 1-T=8.2(3)\%$ for the largest magnetic field applied ($144\,\mu$T). 
As the magnetic field strength is reduced, the spectrum shows a lower peak extinction and multiple, broad resonances. 
This is in stark contrast to the strong extinction~($\approx22\%$) we observed in our previous experiments with the same optical setup but with a circularly polarized FORT~\cite{Chin2017a}. 
We then repeat the experiment, but this time, after polarization gradient cooling, we lower the trap depth to 0.88\,mK. 
This increases the observed extinction significantly [Fig.\ref{fig:tx_matrix}(c)]. 
For our largest magnetic field the transmission spectrum consists of a single Lorentzian line with high peak extinction~$\epsilon=23.3(3)\%$.  

To better understand the effect of the tensor light shift on the transmission spectrum, we numerically calculate the dynamics for the 12-level system containing the $F=2$ and $F'=3$ manifolds. 
Aside from the Zeeman and light shifts [Eq.~(\ref{eq:zeeman}-\ref{eq:Hex})], we include a term in the Hamiltonian that
describes the interaction with the $\sigma^-$-polarized probe light detuned
from the natural $F=2$ to $F'=3$ transition frequency $\omega_{0}$ by $\Delta
= \omega_{p}-\omega_{0}$,
\begin{equation}\hat{H}_\textrm{int}=-\frac{\hbar}{2}   \Omega \hat{A}_- + \text{h.c.}\,,
\end{equation}
where $\hat{A}_-$  is the atomic lowering operator. For the total Hamiltonian
\begin{eqnarray}\label{eq:totalhamiltonian}
\hat{H}&=&  \hat{H}_{0} + \hat{H}_{\textrm{B},F=2} +
\hat{H}_{\textrm{B},F'=3}\nonumber\\
&&{} + \hat{H}_{\textrm{LS},F=2} + \hat{H}_{\textrm{LS},F'=3} + \hat{H}_{\textrm{int}}\,,
\end{eqnarray}
with $\hat{H}_{0}= -\Delta \mathbb{1}_{F=2}$,  where $\mathbb{1}_{F=2}$ is the
unity operator acting on the $F=2$ manifold, we numerically solve the
corresponding master equation
\begin{equation}\dot{\rho}=-\frac{i}{\hbar}[\rho,\hat{H}] + \mathcal{L}[\rho]\,,
\end{equation}
with a Lindblad superoperator $\mathcal{L}[\rho]$ to account for
spontaneous emission.

We initialize in $\ket{\uparrow}$ and apply the probe field for a time~$\tau=1\,\textrm{ms}\gg1/\Omega \gg1/\Gamma$. 
Comparing the number of scattered photons during the probe phase,
\begin{equation}
n_p(\Delta)=\int^{\tau}_0 \textrm{Tr}\left(\rho(t) P_{F'=3} \right) \Gamma dt\,,
\end{equation}
with the value expected for a resonantly driven two-level system,
\begin{equation}
n_{2l}=\frac{\Omega^2}{\Gamma^2+2\Omega^2} \Gamma \tau \approx100\,,
\end{equation}
we obtain an
expected reduction $\eta=n_p(\Delta)/n_{2l}$ of the absorption. Here, $P_{F'=3}$ is the projector on the $F'=3$ manifold. 
The estimated transmission as a function of probe detuning is then
\begin{equation}T(\Delta) = 1- \epsilon_0 \eta(\Delta)\,,
\end{equation}
where $\epsilon_0$ is the resonant, two-level extinction value which depends on the spatial mode of the probe fields.

\begin{figure} 
\centering
  \includegraphics[width=\columnwidth]{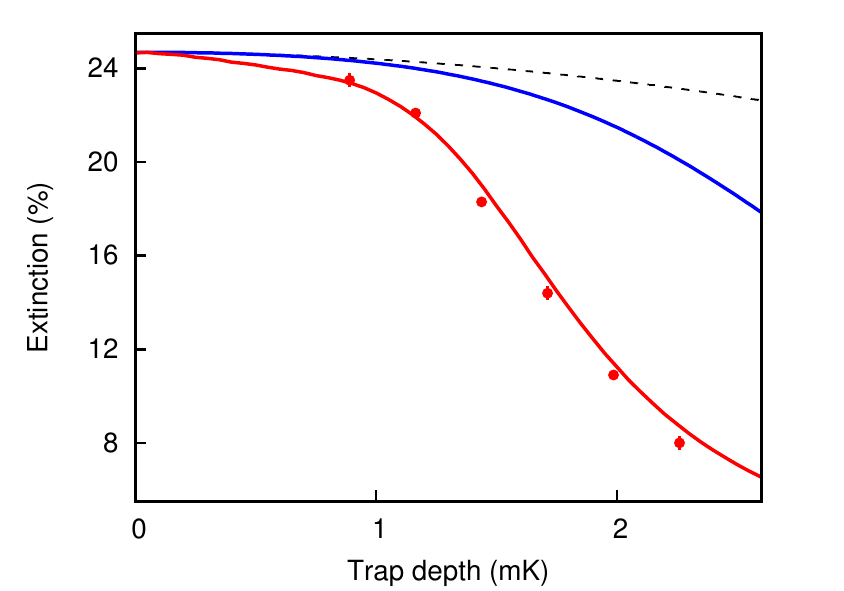}
  \caption{\label{fig:tx_over_trapdepth} Resonant extinction for various trap depths. Red circles: measured extinction, red line: full 12-level numerical simulation,  black dashed line: first order approximation $\epsilon_0 \eta_\textrm{dme}^{(1)}$~[Eq.~(\ref{eq:eta})], blue line: numerically calculated~$\epsilon_0\eta_\textrm{dme}$. }
\end{figure}
We find an excellent agreement between the observed spectrum $T(\Delta)$ and the model for $\epsilon_0=24.7\%$, a value consistent with our previous experiment~\cite{Chin2017a}~[Fig.~\ref{fig:tx_matrix}(b,d)]. 
We further test our model by comparing the resonant extinction at various trap depths but a fixed magnetic field strength of $144\,\mu$T~(Fig.~\ref{fig:tx_over_trapdepth}). 
Again the model matches the experimental data well. 
To further understand the scattering process, 
we consider the relevant dipole matrix element: 
For vanishing light shift both, $\ket{\uparrow}$  and $\ket{\textrm{aux}}$, are energy eigenstates. 
The $\sigma^-$-polarized probe beam couples these two states with a dipole matrix element $|\braket{\uparrow|\epsilon_{\sigma^-}\textbf{d}| \textrm{aux}}|^2$, where the $\epsilon_{\sigma^-}$ is the polarization vector of the probe beam, $\textbf{d}$ is the electric dipole operator.  
With increasing transverse tensor shift~[Eq.~(\ref{eq:Hex})], the angular
momentum eigenstate $\ket{\textrm{aux}}$ is no longer an energy eigenstate --
the corresponding eigenstate~$\ket{\widetilde{\textrm{aux}}}$ of
Eq.~(\ref{eq:totalhamiltonian}) gets admixtures from other $m_z$-states of the same level. 
Thus, the optical coupling strength is reduced by the relative reduction of the dipole matrix element
\begin{equation}
\eta_\textrm{dme}=\frac{ |\braket{\uparrow|\epsilon_{\sigma^-}\textbf{d}| \widetilde{\textrm{aux}}}|^2}{|\braket{\uparrow|\epsilon_{\sigma^-}\textbf{d}| \textrm{aux}}|^2}.
\end{equation}
For strong magnetic fields, $\hbar \omega_L \gg U_0 c_{t}$, the reduction of the dipole matrix element
in first order approximation is given by 
\begin{equation}\label{eq:eta}
 \eta_\textrm{dme}^{(1)} = 1 -\frac{15}{8} \left(\frac{U_0 c_{t}}{\hbar\omega_L}\right)^2\,.
\end{equation}
However, neither $\epsilon_0 \eta_\textrm{dme}^{(1)}$ nor the numerically calculated~$\epsilon_0\eta_\textrm{dme}$ reproduces our measured values well~(Fig.~\ref{fig:tx_over_trapdepth}). 
The reason is that the observed spectrum is strongly affected by multiple scattering events. 
When the energy eigenstates are superpositions of $m_z$ states, there is a probability that a scattering event brings the atom out of the $\left\{\ket{\uparrow}, \ket{\textrm{aux}}\right\}$ subspace. 
After such an optical depolarization event, the resonance frequency is shifted, and thus the optical coupling is strongly reduced.
The full numerical simulation takes these spin flips into account, resulting in a good match with the experimental data. 

From the comparison between experiment and theory we learn that (I) it is indeed the tensor light shift that causes the complexity of the transmission spectrum, 
(II) the spin dynamics induced by multiple scattering events are important for the spectral shape of the spectrum, 
and (III) the optical coupling between $\ket{\uparrow}$ and $\ket{\textrm{aux}}$ is close to an ideal two-level system in a shallow trap with a strong magnetic field applied. 

\section{Ground State Qubit}
We characterize the ground state qubit in terms of state readout fidelity and
coherence time to show that efficient optical coupling and a long coherence
time of the qubit can be simultaneously achieved. 
For the following experiments, we choose a trap depth $U_0=k_B\times 0.88\,$mK
and a magnetic field strength of $144\,\mu$T, in which the highest optical coupling is observed. 
The state readout fidelity is determined by preparing the atom in a particular state, and then illuminating the atom for $600\,\mu$s with light resonant with the 5\lev{S}{1}{2}  $F=2$ to 5\lev{P}{3}{2}  $F'=3$ transition~\cite{Fuhrmanek2011,Gibbons2011}. 
The number of photons detected at $D_1$ and $D_2$ allows us to infer the qubit state. 
When the atom is initially in $\ket{\uparrow}$, we detect a mean number of $n_\uparrow = 9.85(8)$~photons. 
For an atom in $\ket{\downarrow}$, the atom ideally scatters almost no photons
because of the large hyperfine splitting of $6.8$\,GHz, but we occasionally 
register one or two detection events~(mean number of $n_\downarrow = 0.17(1)$~photons).
For this measurement, we indiscriminately prepare the atom in the
5\lev{S}{1}{2} $F=1$  as the same dark state behavior is expected for all three Zeeman levels. 
\begin{figure} 
\centering
  \includegraphics[width=\columnwidth]{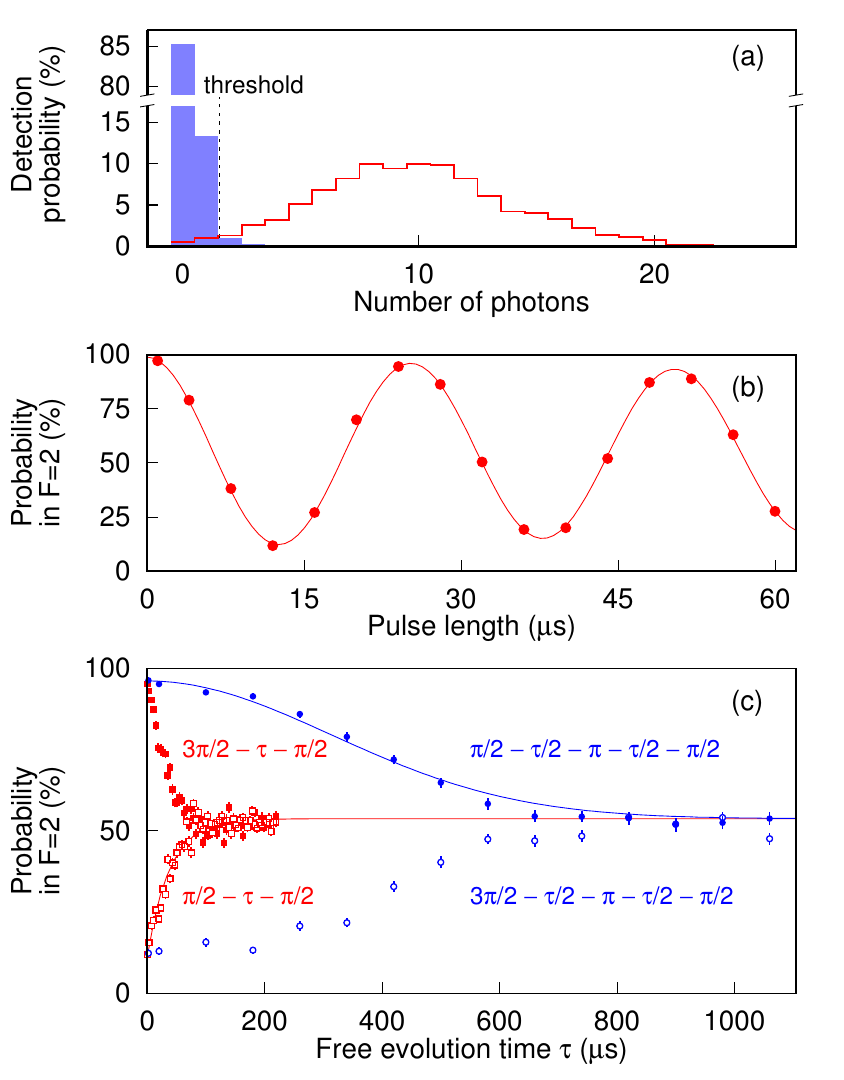}
  \caption{\label{fig:statereadout_histo}
  (a)~Histogram of photon detection probability for atoms prepared in $F=1$
  (blue) and $F=2$ (red), respectively. 
  (b)~Rabi oscillation between $\ket{\uparrow}$ and $\ket{\downarrow}$. 
  (c)~Ramsey (red) and spin-echo (blue) when the atom is
  initially prepared in $\ket{\uparrow}$~(solid symbols) or
  $\ket{\downarrow}$~(empty symbols). We fit a decaying exponential to the
  Ramsey signal and a decaying Gaussian to the spin-echo signal to extract the
  $1/e$ time constants, $T^*_2= 38(3)\mu$s and $T_2=446(14)\,\mu$s.  }
\end{figure}

Figure~\ref{fig:statereadout_histo}(a) shows the histogram of $n_\uparrow$ and $n_\downarrow$ after 3000 repetitions of the experiment. From the histogram we determine a threshold value~$n_\textrm{th}=2$ that optimizes the discrimination between the two distributions. The probabilities for erroneous state assignment are $\chi_\uparrow =1.5\%$ and $\chi_\downarrow =1.4\%$ for an atom prepared in $\ket{\uparrow}$ and $\ket{\downarrow}$, respectively. 
Thus, we achieve a state readout fidelity $F= 1 - (\chi_\uparrow + \chi_\downarrow)/2=98.6(2)\%$, similar to previously reported values~\cite{Fuhrmanek2011,Gibbons2011,Kwon2017,Martinez-Dorantes2017}. 

To characterize the qubit coherence properties, we apply a microwave field resonant with the $\ket{\uparrow}$ to $\ket{\downarrow}$ transition to drive Rabi oscillations and perform Ramsey and spin-echo sequences~\cite{Kuhr2003,Andersen2003,Andersen2004,Kuhr2005,Rosenfeld2011}.  
We observe a Rabi frequency of $\Omega_{\textrm{mw}}=2\pi\times 39.6(5)\,$kHz
with a visibility of $0.89(1)$ and little damping in the first $60\,\mu$s [Fig.~\ref{fig:statereadout_histo}(b)]. 
The dephasing time is determined from a Ramsey experiment, where we apply two resonant microwave pulses for $t_{\pi/2}=\pi/(2 \Omega)$ separated by a free evolution time $\tau$~[Fig.~\ref{fig:statereadout_histo}(c)]. 
We fit an exponential decay to the Ramsey contrast, and extract the dephasing time~$T^*_2= 38(3)\mu$s. 
Subsequently, applying a spin-echo, i.e. inserting an extra microwave pulse for
$t_{\pi}=\pi / \Omega$ halfway in the free evolution period~$\tau$, the inhomogeneous dephasing is reversed and we observe a much slower decay of the contrast. 
Fitting to a decaying Gaussian, we obtain the coherence time~$T_2=446(14)\,\mu$s defined as the $1/e$ decay time of the spin-echo visibility. 
Despite not using the clock states, we achieve a high ratio of state manipulation speed and preserved coherence, $T_2/t_{\pi}\approx 35$. 
The Rabi frequency can be further increased by using fast optical Raman transitions~\cite{Yavuz2006,Jones2007}.
Figure~\ref{fig:statereadout_histo}(c) also displays the results of
Ramsey and spin-echo experiments where we initially prepare the atom in state
$\ket{\downarrow}$. 
As expected, the observed values mirror the experiments starting from~$\ket{\uparrow}$.

\section{Conclusion}
We have shown that a combination of a shallow optical trap and strong magnetic fields sufficiently mitigates the effects of the tensor light shift on the optical coupling. 
Under these conditions, we demonstrated high qubit coherence and readout fidelity.
The capability of coupling qubit states selectively to well-defined optical channels enables new ways of building up hybrid light-atom quantum states.
In particular, we expect that several protocols that were originally developed
for solid state quantum systems -- where qubit state-selective, closed optical transitions are common -- can be realized with a neutral atom in a
dipole trap. This includes the generation of time-bin atom-photon entanglement~\cite{Lee2018} and the sequential generation of entangled photons~\cite{Lindner2009,Schwartz2016}.

\begin{acknowledgments}
We acknowledge the support of this work by the Ministry of Education in
Singapore (AcRF Tier 1) and the National Research Foundation, Prime Minister's
office.
M.\,Steiner acknowledges support by the Lee Kuan Yew Postdoctoral Fellowship.
\end{acknowledgments}

\bibliographystyle{apsrev4-1}

\begin{thebibliography}{28}%
\makeatletter
\providecommand \@ifxundefined [1]{%
 \@ifx{#1\undefined}
}%
\providecommand \@ifnum [1]{%
 \ifnum #1\expandafter \@firstoftwo
 \else \expandafter \@secondoftwo
 \fi
}%
\providecommand \@ifx [1]{%
 \ifx #1\expandafter \@firstoftwo
 \else \expandafter \@secondoftwo
 \fi
}%
\providecommand \natexlab [1]{#1}%
\providecommand \enquote  [1]{``#1''}%
\providecommand \bibnamefont  [1]{#1}%
\providecommand \bibfnamefont [1]{#1}%
\providecommand \citenamefont [1]{#1}%
\providecommand \href@noop [0]{\@secondoftwo}%
\providecommand \href [0]{\begingroup \@sanitize@url \@href}%
\providecommand \@href[1]{\@@startlink{#1}\@@href}%
\providecommand \@@href[1]{\endgroup#1\@@endlink}%
\providecommand \@sanitize@url [0]{\catcode `\\12\catcode `\$12\catcode
  `\&12\catcode `\#12\catcode `\^12\catcode `\_12\catcode `\%12\relax}%
\providecommand \@@startlink[1]{}%
\providecommand \@@endlink[0]{}%
\providecommand \url  [0]{\begingroup\@sanitize@url \@url }%
\providecommand \@url [1]{\endgroup\@href {#1}{\urlprefix }}%
\providecommand \urlprefix  [0]{URL }%
\providecommand \Eprint [0]{\href }%
\providecommand \doibase [0]{http://dx.doi.org/}%
\providecommand \selectlanguage [0]{\@gobble}%
\providecommand \bibinfo  [0]{\@secondoftwo}%
\providecommand \bibfield  [0]{\@secondoftwo}%
\providecommand \translation [1]{[#1]}%
\providecommand \BibitemOpen [0]{}%
\providecommand \bibitemStop [0]{}%
\providecommand \bibitemNoStop [0]{.\EOS\space}%
\providecommand \EOS [0]{\spacefactor3000\relax}%
\providecommand \BibitemShut  [1]{\csname bibitem#1\endcsname}%
\let\auto@bib@innerbib\@empty
\bibitem [{\citenamefont {Bernien}\ \emph {et~al.}(2017)\citenamefont
  {Bernien}, \citenamefont {Schwartz}, \citenamefont {Keesling}, \citenamefont
  {Levine}, \citenamefont {Omran}, \citenamefont {Pichler}, \citenamefont
  {Choi}, \citenamefont {Zibrov}, \citenamefont {Endres}, \citenamefont
  {Greiner}, \citenamefont {Vuleti\'{c}},\ and\ \citenamefont
  {Lukin}}]{Bernien2017}%
  \BibitemOpen
  \bibfield  {author} {\bibinfo {author} {\bibfnamefont {H.}~\bibnamefont
  {Bernien}}, \bibinfo {author} {\bibfnamefont {S.}~\bibnamefont {Schwartz}},
  \bibinfo {author} {\bibfnamefont {A.}~\bibnamefont {Keesling}}, \bibinfo
  {author} {\bibfnamefont {H.}~\bibnamefont {Levine}}, \bibinfo {author}
  {\bibfnamefont {A.}~\bibnamefont {Omran}}, \bibinfo {author} {\bibfnamefont
  {H.}~\bibnamefont {Pichler}}, \bibinfo {author} {\bibfnamefont
  {S.}~\bibnamefont {Choi}}, \bibinfo {author} {\bibfnamefont {A.~S.}\
  \bibnamefont {Zibrov}}, \bibinfo {author} {\bibfnamefont {M.}~\bibnamefont
  {Endres}}, \bibinfo {author} {\bibfnamefont {M.}~\bibnamefont {Greiner}},
  \bibinfo {author} {\bibfnamefont {V.}~\bibnamefont {Vuleti\'{c}}}, \ and\
  \bibinfo {author} {\bibfnamefont {M.~D.}\ \bibnamefont {Lukin}},\ }\href
  {http://dx.doi.org/10.1038/nature24622} {\bibfield  {journal} {\bibinfo
  {journal} {Nature}\ }\textbf {\bibinfo {volume} {551}},\ \bibinfo {pages}
  {579} (\bibinfo {year} {2017})}\BibitemShut {NoStop}%
\bibitem [{\citenamefont {Lienhard}\ \emph {et~al.}(2018)\citenamefont
  {Lienhard}, \citenamefont {de~L\'es\'eleuc}, \citenamefont {Barredo},
  \citenamefont {Lahaye}, \citenamefont {Browaeys}, \citenamefont {Schuler},
  \citenamefont {Henry},\ and\ \citenamefont {L\"auchli}}]{Lienhard2018}%
  \BibitemOpen
  \bibfield  {author} {\bibinfo {author} {\bibfnamefont {V.}~\bibnamefont
  {Lienhard}}, \bibinfo {author} {\bibfnamefont {S.}~\bibnamefont
  {de~L\'es\'eleuc}}, \bibinfo {author} {\bibfnamefont {D.}~\bibnamefont
  {Barredo}}, \bibinfo {author} {\bibfnamefont {T.}~\bibnamefont {Lahaye}},
  \bibinfo {author} {\bibfnamefont {A.}~\bibnamefont {Browaeys}}, \bibinfo
  {author} {\bibfnamefont {M.}~\bibnamefont {Schuler}}, \bibinfo {author}
  {\bibfnamefont {L.-P.}\ \bibnamefont {Henry}}, \ and\ \bibinfo {author}
  {\bibfnamefont {A.~M.}\ \bibnamefont {L\"auchli}},\ }\href {\doibase
  10.1103/PhysRevX.8.021070} {\bibfield  {journal} {\bibinfo  {journal} {Phys.
  Rev. X}\ }\textbf {\bibinfo {volume} {8}},\ \bibinfo {pages} {021070}
  (\bibinfo {year} {2018})}\BibitemShut {NoStop}%
\bibitem [{\citenamefont {Wilk}\ \emph {et~al.}(2010)\citenamefont {Wilk},
  \citenamefont {Ga\"etan}, \citenamefont {Evellin}, \citenamefont {Wolters},
  \citenamefont {Miroshnychenko}, \citenamefont {Grangier},\ and\ \citenamefont
  {Browaeys}}]{Wilk2010}%
  \BibitemOpen
  \bibfield  {author} {\bibinfo {author} {\bibfnamefont {T.}~\bibnamefont
  {Wilk}}, \bibinfo {author} {\bibfnamefont {A.}~\bibnamefont {Ga\"etan}},
  \bibinfo {author} {\bibfnamefont {C.}~\bibnamefont {Evellin}}, \bibinfo
  {author} {\bibfnamefont {J.}~\bibnamefont {Wolters}}, \bibinfo {author}
  {\bibfnamefont {Y.}~\bibnamefont {Miroshnychenko}}, \bibinfo {author}
  {\bibfnamefont {P.}~\bibnamefont {Grangier}}, \ and\ \bibinfo {author}
  {\bibfnamefont {A.}~\bibnamefont {Browaeys}},\ }\href {\doibase
  10.1103/PhysRevLett.104.010502} {\bibfield  {journal} {\bibinfo  {journal}
  {Phys. Rev. Lett.}\ }\textbf {\bibinfo {volume} {104}},\ \bibinfo {pages}
  {010502} (\bibinfo {year} {2010})}\BibitemShut {NoStop}%
\bibitem [{\citenamefont {Ritter}\ \emph {et~al.}(2012)\citenamefont {Ritter},
  \citenamefont {Nölleke}, \citenamefont {Hahn}, \citenamefont {Reiserer},
  \citenamefont {Neuzner}, \citenamefont {Uphoff}, \citenamefont {Mücke},
  \citenamefont {Figueroa}, \citenamefont {Bochmann},\ and\ \citenamefont
  {Rempe}}]{Ritter2012}%
  \BibitemOpen
  \bibfield  {author} {\bibinfo {author} {\bibfnamefont {S.}~\bibnamefont
  {Ritter}}, \bibinfo {author} {\bibfnamefont {C.}~\bibnamefont {Nölleke}},
  \bibinfo {author} {\bibfnamefont {C.}~\bibnamefont {Hahn}}, \bibinfo {author}
  {\bibfnamefont {A.}~\bibnamefont {Reiserer}}, \bibinfo {author}
  {\bibfnamefont {A.}~\bibnamefont {Neuzner}}, \bibinfo {author} {\bibfnamefont
  {M.}~\bibnamefont {Uphoff}}, \bibinfo {author} {\bibfnamefont
  {M.}~\bibnamefont {Mücke}}, \bibinfo {author} {\bibfnamefont
  {E.}~\bibnamefont {Figueroa}}, \bibinfo {author} {\bibfnamefont
  {J.}~\bibnamefont {Bochmann}}, \ and\ \bibinfo {author} {\bibfnamefont
  {G.}~\bibnamefont {Rempe}},\ }\href {http://dx.doi.org/10.1038/nature11023}
  {\bibfield  {journal} {\bibinfo  {journal} {Nature}\ }\textbf {\bibinfo
  {volume} {484}},\ \bibinfo {pages} {195} (\bibinfo {year}
  {2012})}\BibitemShut {NoStop}%
\bibitem [{\citenamefont {Tey}\ \emph {et~al.}(2008)\citenamefont {Tey},
  \citenamefont {Chen}, \citenamefont {Aljunid}, \citenamefont {Chng},
  \citenamefont {Huber}, \citenamefont {Maslennikov},\ and\ \citenamefont
  {Kurtsiefer}}]{Tey2008}%
  \BibitemOpen
  \bibfield  {author} {\bibinfo {author} {\bibfnamefont {M.~K.}\ \bibnamefont
  {Tey}}, \bibinfo {author} {\bibfnamefont {Z.}~\bibnamefont {Chen}}, \bibinfo
  {author} {\bibfnamefont {S.~A.}\ \bibnamefont {Aljunid}}, \bibinfo {author}
  {\bibfnamefont {B.}~\bibnamefont {Chng}}, \bibinfo {author} {\bibfnamefont
  {F.}~\bibnamefont {Huber}}, \bibinfo {author} {\bibfnamefont
  {G.}~\bibnamefont {Maslennikov}}, \ and\ \bibinfo {author} {\bibfnamefont
  {C.}~\bibnamefont {Kurtsiefer}},\ }\href
  {http://dx.doi.org/10.1038/nphys1096} {\bibfield  {journal} {\bibinfo
  {journal} {Nat Phys}\ }\textbf {\bibinfo {volume} {4}},\ \bibinfo {pages}
  {924} (\bibinfo {year} {2008})}\BibitemShut {NoStop}%
\bibitem [{\citenamefont {Chin}\ \emph
  {et~al.}(2017{\natexlab{a}})\citenamefont {Chin}, \citenamefont {Steiner},\
  and\ \citenamefont {Kurtsiefer}}]{Chin2017a}%
  \BibitemOpen
  \bibfield  {author} {\bibinfo {author} {\bibfnamefont {Y.-S.}\ \bibnamefont
  {Chin}}, \bibinfo {author} {\bibfnamefont {M.}~\bibnamefont {Steiner}}, \
  and\ \bibinfo {author} {\bibfnamefont {C.}~\bibnamefont {Kurtsiefer}},\
  }\href {https://doi.org/10.1038/s41467-017-01495-3} {\bibfield  {journal}
  {\bibinfo  {journal} {Nature Communications}\ }\textbf {\bibinfo {volume}
  {8}},\ \bibinfo {pages} {1200} (\bibinfo {year}
  {2017}{\natexlab{a}})}\BibitemShut {NoStop}%
\bibitem [{\citenamefont {Aljunid}\ \emph {et~al.}(2013)\citenamefont
  {Aljunid}, \citenamefont {Maslennikov}, \citenamefont {Wang}, \citenamefont
  {Dao}, \citenamefont {Scarani},\ and\ \citenamefont
  {Kurtsiefer}}]{Aljunid2013}%
  \BibitemOpen
  \bibfield  {author} {\bibinfo {author} {\bibfnamefont {S.~A.}\ \bibnamefont
  {Aljunid}}, \bibinfo {author} {\bibfnamefont {G.}~\bibnamefont
  {Maslennikov}}, \bibinfo {author} {\bibfnamefont {Y.}~\bibnamefont {Wang}},
  \bibinfo {author} {\bibfnamefont {H.~L.}\ \bibnamefont {Dao}}, \bibinfo
  {author} {\bibfnamefont {V.}~\bibnamefont {Scarani}}, \ and\ \bibinfo
  {author} {\bibfnamefont {C.}~\bibnamefont {Kurtsiefer}},\ }\href {\doibase
  10.1103/PhysRevLett.111.103001} {\bibfield  {journal} {\bibinfo  {journal}
  {Phys. Rev. Lett.}\ }\textbf {\bibinfo {volume} {111}},\ \bibinfo {pages}
  {103001} (\bibinfo {year} {2013})}\BibitemShut {NoStop}%
\bibitem [{\citenamefont {Leong}\ \emph {et~al.}(2016)\citenamefont {Leong},
  \citenamefont {Seidler}, \citenamefont {Steiner}, \citenamefont {Cerè},\
  and\ \citenamefont {Kurtsiefer}}]{Leong2016}%
  \BibitemOpen
  \bibfield  {author} {\bibinfo {author} {\bibfnamefont {V.}~\bibnamefont
  {Leong}}, \bibinfo {author} {\bibfnamefont {M.~A.}\ \bibnamefont {Seidler}},
  \bibinfo {author} {\bibfnamefont {M.}~\bibnamefont {Steiner}}, \bibinfo
  {author} {\bibfnamefont {A.}~\bibnamefont {Cerè}}, \ and\ \bibinfo {author}
  {\bibfnamefont {C.}~\bibnamefont {Kurtsiefer}},\ }\href
  {http://dx.doi.org/10.1038/ncomms13716} {\bibfield  {journal} {\bibinfo
  {journal} {Nature Communications}\ }\textbf {\bibinfo {volume} {7}},\
  \bibinfo {pages} {13716} (\bibinfo {year} {2016})}\BibitemShut {NoStop}%
\bibitem [{\citenamefont {Steiner}\ \emph {et~al.}(2017)\citenamefont
  {Steiner}, \citenamefont {Leong}, \citenamefont {Seidler}, \citenamefont
  {Cer\`{e}},\ and\ \citenamefont {Kurtsiefer}}]{Steiner2017}%
  \BibitemOpen
  \bibfield  {author} {\bibinfo {author} {\bibfnamefont {M.}~\bibnamefont
  {Steiner}}, \bibinfo {author} {\bibfnamefont {V.}~\bibnamefont {Leong}},
  \bibinfo {author} {\bibfnamefont {M.~A.}\ \bibnamefont {Seidler}}, \bibinfo
  {author} {\bibfnamefont {A.}~\bibnamefont {Cer\`{e}}}, \ and\ \bibinfo
  {author} {\bibfnamefont {C.}~\bibnamefont {Kurtsiefer}},\ }\href {\doibase
  10.1364/OE.25.006294} {\bibfield  {journal} {\bibinfo  {journal} {Opt.
  Express}\ }\textbf {\bibinfo {volume} {25}},\ \bibinfo {pages} {6294}
  (\bibinfo {year} {2017})}\BibitemShut {NoStop}%
\bibitem [{\citenamefont {Deutsch}\ and\ \citenamefont
  {Jessen}(2010)}]{Deutsch2010}%
  \BibitemOpen
  \bibfield  {author} {\bibinfo {author} {\bibfnamefont {I.~H.}\ \bibnamefont
  {Deutsch}}\ and\ \bibinfo {author} {\bibfnamefont {P.~S.}\ \bibnamefont
  {Jessen}},\ }\href {\doibase https://doi.org/10.1016/j.optcom.2009.10.059}
  {\bibfield  {journal} {\bibinfo  {journal} {Optics Communications}\ }\textbf
  {\bibinfo {volume} {283}},\ \bibinfo {pages} {681 } (\bibinfo {year}
  {2010})}\BibitemShut {NoStop}%
\bibitem [{\citenamefont {Romalis}\ and\ \citenamefont
  {Fortson}(1999)}]{Romalis1999}%
  \BibitemOpen
  \bibfield  {author} {\bibinfo {author} {\bibfnamefont {M.~V.}\ \bibnamefont
  {Romalis}}\ and\ \bibinfo {author} {\bibfnamefont {E.~N.}\ \bibnamefont
  {Fortson}},\ }\href {\doibase 10.1103/PhysRevA.59.4547} {\bibfield  {journal}
  {\bibinfo  {journal} {Phys. Rev. A}\ }\textbf {\bibinfo {volume} {59}},\
  \bibinfo {pages} {4547} (\bibinfo {year} {1999})}\BibitemShut {NoStop}%
\bibitem [{\citenamefont {Schlosser}\ \emph {et~al.}(2001)\citenamefont
  {Schlosser}, \citenamefont {Reymond}, \citenamefont {Protsenko},\ and\
  \citenamefont {Grangier}}]{Schlosser2001}%
  \BibitemOpen
  \bibfield  {author} {\bibinfo {author} {\bibfnamefont {N.}~\bibnamefont
  {Schlosser}}, \bibinfo {author} {\bibfnamefont {G.}~\bibnamefont {Reymond}},
  \bibinfo {author} {\bibfnamefont {I.}~\bibnamefont {Protsenko}}, \ and\
  \bibinfo {author} {\bibfnamefont {P.}~\bibnamefont {Grangier}},\ }\href
  {http://dx.doi.org/10.1038/35082512} {\bibfield  {journal} {\bibinfo
  {journal} {Nature}\ }\textbf {\bibinfo {volume} {411}},\ \bibinfo {pages}
  {1024} (\bibinfo {year} {2001})}\BibitemShut {NoStop}%
\bibitem [{\citenamefont {Chin}\ \emph
  {et~al.}(2017{\natexlab{b}})\citenamefont {Chin}, \citenamefont {Steiner},\
  and\ \citenamefont {Kurtsiefer}}]{Chin2017}%
  \BibitemOpen
  \bibfield  {author} {\bibinfo {author} {\bibfnamefont {Y.-S.}\ \bibnamefont
  {Chin}}, \bibinfo {author} {\bibfnamefont {M.}~\bibnamefont {Steiner}}, \
  and\ \bibinfo {author} {\bibfnamefont {C.}~\bibnamefont {Kurtsiefer}},\
  }\href {\doibase 10.1103/PhysRevA.95.043809} {\bibfield  {journal} {\bibinfo
  {journal} {Phys. Rev. A}\ }\textbf {\bibinfo {volume} {95}},\ \bibinfo
  {pages} {043809} (\bibinfo {year} {2017}{\natexlab{b}})}\BibitemShut
  {NoStop}%
\bibitem [{\citenamefont {Chin}\ \emph
  {et~al.}(2017{\natexlab{c}})\citenamefont {Chin}, \citenamefont {Steiner},\
  and\ \citenamefont {Kurtsiefer}}]{Chin2017b}%
  \BibitemOpen
  \bibfield  {author} {\bibinfo {author} {\bibfnamefont {Y.-S.}\ \bibnamefont
  {Chin}}, \bibinfo {author} {\bibfnamefont {M.}~\bibnamefont {Steiner}}, \
  and\ \bibinfo {author} {\bibfnamefont {C.}~\bibnamefont {Kurtsiefer}},\
  }\href {\doibase 10.1103/PhysRevA.96.033406} {\bibfield  {journal} {\bibinfo
  {journal} {Phys. Rev. A}\ }\textbf {\bibinfo {volume} {96}},\ \bibinfo
  {pages} {033406} (\bibinfo {year} {2017}{\natexlab{c}})}\BibitemShut
  {NoStop}%
\bibitem [{\citenamefont {Fuhrmanek}\ \emph {et~al.}(2011)\citenamefont
  {Fuhrmanek}, \citenamefont {Bourgain}, \citenamefont {Sortais},\ and\
  \citenamefont {Browaeys}}]{Fuhrmanek2011}%
  \BibitemOpen
  \bibfield  {author} {\bibinfo {author} {\bibfnamefont {A.}~\bibnamefont
  {Fuhrmanek}}, \bibinfo {author} {\bibfnamefont {R.}~\bibnamefont {Bourgain}},
  \bibinfo {author} {\bibfnamefont {Y.~R.~P.}\ \bibnamefont {Sortais}}, \ and\
  \bibinfo {author} {\bibfnamefont {A.}~\bibnamefont {Browaeys}},\ }\href
  {\doibase 10.1103/PhysRevLett.106.133003} {\bibfield  {journal} {\bibinfo
  {journal} {Phys. Rev. Lett.}\ }\textbf {\bibinfo {volume} {106}},\ \bibinfo
  {pages} {133003} (\bibinfo {year} {2011})}\BibitemShut {NoStop}%
\bibitem [{\citenamefont {Gibbons}\ \emph {et~al.}(2011)\citenamefont
  {Gibbons}, \citenamefont {Hamley}, \citenamefont {Shih},\ and\ \citenamefont
  {Chapman}}]{Gibbons2011}%
  \BibitemOpen
  \bibfield  {author} {\bibinfo {author} {\bibfnamefont {M.~J.}\ \bibnamefont
  {Gibbons}}, \bibinfo {author} {\bibfnamefont {C.~D.}\ \bibnamefont {Hamley}},
  \bibinfo {author} {\bibfnamefont {C.-Y.}\ \bibnamefont {Shih}}, \ and\
  \bibinfo {author} {\bibfnamefont {M.~S.}\ \bibnamefont {Chapman}},\ }\href
  {\doibase 10.1103/PhysRevLett.106.133002} {\bibfield  {journal} {\bibinfo
  {journal} {Phys. Rev. Lett.}\ }\textbf {\bibinfo {volume} {106}},\ \bibinfo
  {pages} {133002} (\bibinfo {year} {2011})}\BibitemShut {NoStop}%
\bibitem [{\citenamefont {Kwon}\ \emph {et~al.}(2017)\citenamefont {Kwon},
  \citenamefont {Ebert}, \citenamefont {Walker},\ and\ \citenamefont
  {Saffman}}]{Kwon2017}%
  \BibitemOpen
  \bibfield  {author} {\bibinfo {author} {\bibfnamefont {M.}~\bibnamefont
  {Kwon}}, \bibinfo {author} {\bibfnamefont {M.~F.}\ \bibnamefont {Ebert}},
  \bibinfo {author} {\bibfnamefont {T.~G.}\ \bibnamefont {Walker}}, \ and\
  \bibinfo {author} {\bibfnamefont {M.}~\bibnamefont {Saffman}},\ }\href
  {\doibase 10.1103/PhysRevLett.119.180504} {\bibfield  {journal} {\bibinfo
  {journal} {Phys. Rev. Lett.}\ }\textbf {\bibinfo {volume} {119}},\ \bibinfo
  {pages} {180504} (\bibinfo {year} {2017})}\BibitemShut {NoStop}%
\bibitem [{\citenamefont {Martinez-Dorantes}\ \emph {et~al.}(2017)\citenamefont
  {Martinez-Dorantes}, \citenamefont {Alt}, \citenamefont {Gallego},
  \citenamefont {Ghosh}, \citenamefont {Ratschbacher}, \citenamefont
  {V\"olzke},\ and\ \citenamefont {Meschede}}]{Martinez-Dorantes2017}%
  \BibitemOpen
  \bibfield  {author} {\bibinfo {author} {\bibfnamefont {M.}~\bibnamefont
  {Martinez-Dorantes}}, \bibinfo {author} {\bibfnamefont {W.}~\bibnamefont
  {Alt}}, \bibinfo {author} {\bibfnamefont {J.}~\bibnamefont {Gallego}},
  \bibinfo {author} {\bibfnamefont {S.}~\bibnamefont {Ghosh}}, \bibinfo
  {author} {\bibfnamefont {L.}~\bibnamefont {Ratschbacher}}, \bibinfo {author}
  {\bibfnamefont {Y.}~\bibnamefont {V\"olzke}}, \ and\ \bibinfo {author}
  {\bibfnamefont {D.}~\bibnamefont {Meschede}},\ }\href {\doibase
  10.1103/PhysRevLett.119.180503} {\bibfield  {journal} {\bibinfo  {journal}
  {Phys. Rev. Lett.}\ }\textbf {\bibinfo {volume} {119}},\ \bibinfo {pages}
  {180503} (\bibinfo {year} {2017})}\BibitemShut {NoStop}%
\bibitem [{\citenamefont {Kuhr}\ \emph {et~al.}(2003)\citenamefont {Kuhr},
  \citenamefont {Alt}, \citenamefont {Schrader}, \citenamefont {Dotsenko},
  \citenamefont {Miroshnychenko}, \citenamefont {Rosenfeld}, \citenamefont
  {Khudaverdyan}, \citenamefont {Gomer}, \citenamefont {Rauschenbeutel},\ and\
  \citenamefont {Meschede}}]{Kuhr2003}%
  \BibitemOpen
  \bibfield  {author} {\bibinfo {author} {\bibfnamefont {S.}~\bibnamefont
  {Kuhr}}, \bibinfo {author} {\bibfnamefont {W.}~\bibnamefont {Alt}}, \bibinfo
  {author} {\bibfnamefont {D.}~\bibnamefont {Schrader}}, \bibinfo {author}
  {\bibfnamefont {I.}~\bibnamefont {Dotsenko}}, \bibinfo {author}
  {\bibfnamefont {Y.}~\bibnamefont {Miroshnychenko}}, \bibinfo {author}
  {\bibfnamefont {W.}~\bibnamefont {Rosenfeld}}, \bibinfo {author}
  {\bibfnamefont {M.}~\bibnamefont {Khudaverdyan}}, \bibinfo {author}
  {\bibfnamefont {V.}~\bibnamefont {Gomer}}, \bibinfo {author} {\bibfnamefont
  {A.}~\bibnamefont {Rauschenbeutel}}, \ and\ \bibinfo {author} {\bibfnamefont
  {D.}~\bibnamefont {Meschede}},\ }\href {\doibase
  10.1103/PhysRevLett.91.213002} {\bibfield  {journal} {\bibinfo  {journal}
  {Phys. Rev. Lett.}\ }\textbf {\bibinfo {volume} {91}},\ \bibinfo {pages}
  {213002} (\bibinfo {year} {2003})}\BibitemShut {NoStop}%
\bibitem [{\citenamefont {Andersen}\ \emph {et~al.}(2003)\citenamefont
  {Andersen}, \citenamefont {Kaplan},\ and\ \citenamefont
  {Davidson}}]{Andersen2003}%
  \BibitemOpen
  \bibfield  {author} {\bibinfo {author} {\bibfnamefont {M.~F.}\ \bibnamefont
  {Andersen}}, \bibinfo {author} {\bibfnamefont {A.}~\bibnamefont {Kaplan}}, \
  and\ \bibinfo {author} {\bibfnamefont {N.}~\bibnamefont {Davidson}},\ }\href
  {\doibase 10.1103/PhysRevLett.90.023001} {\bibfield  {journal} {\bibinfo
  {journal} {Phys. Rev. Lett.}\ }\textbf {\bibinfo {volume} {90}},\ \bibinfo
  {pages} {023001} (\bibinfo {year} {2003})}\BibitemShut {NoStop}%
\bibitem [{\citenamefont {Andersen}\ \emph {et~al.}(2004)\citenamefont
  {Andersen}, \citenamefont {Kaplan}, \citenamefont {Gr\"unzweig},\ and\
  \citenamefont {Davidson}}]{Andersen2004}%
  \BibitemOpen
  \bibfield  {author} {\bibinfo {author} {\bibfnamefont {M.~F.}\ \bibnamefont
  {Andersen}}, \bibinfo {author} {\bibfnamefont {A.}~\bibnamefont {Kaplan}},
  \bibinfo {author} {\bibfnamefont {T.}~\bibnamefont {Gr\"unzweig}}, \ and\
  \bibinfo {author} {\bibfnamefont {N.}~\bibnamefont {Davidson}},\ }\href
  {\doibase 10.1103/PhysRevA.70.013405} {\bibfield  {journal} {\bibinfo
  {journal} {Phys. Rev. A}\ }\textbf {\bibinfo {volume} {70}},\ \bibinfo
  {pages} {013405} (\bibinfo {year} {2004})}\BibitemShut {NoStop}%
\bibitem [{\citenamefont {Kuhr}\ \emph {et~al.}(2005)\citenamefont {Kuhr},
  \citenamefont {Alt}, \citenamefont {Schrader}, \citenamefont {Dotsenko},
  \citenamefont {Miroshnychenko}, \citenamefont {Rauschenbeutel},\ and\
  \citenamefont {Meschede}}]{Kuhr2005}%
  \BibitemOpen
  \bibfield  {author} {\bibinfo {author} {\bibfnamefont {S.}~\bibnamefont
  {Kuhr}}, \bibinfo {author} {\bibfnamefont {W.}~\bibnamefont {Alt}}, \bibinfo
  {author} {\bibfnamefont {D.}~\bibnamefont {Schrader}}, \bibinfo {author}
  {\bibfnamefont {I.}~\bibnamefont {Dotsenko}}, \bibinfo {author}
  {\bibfnamefont {Y.}~\bibnamefont {Miroshnychenko}}, \bibinfo {author}
  {\bibfnamefont {A.}~\bibnamefont {Rauschenbeutel}}, \ and\ \bibinfo {author}
  {\bibfnamefont {D.}~\bibnamefont {Meschede}},\ }\href {\doibase
  10.1103/PhysRevA.72.023406} {\bibfield  {journal} {\bibinfo  {journal} {Phys.
  Rev. A}\ }\textbf {\bibinfo {volume} {72}},\ \bibinfo {pages} {023406}
  (\bibinfo {year} {2005})}\BibitemShut {NoStop}%
\bibitem [{\citenamefont {Rosenfeld}\ \emph {et~al.}(2011)\citenamefont
  {Rosenfeld}, \citenamefont {Volz}, \citenamefont {Weber},\ and\ \citenamefont
  {Weinfurter}}]{Rosenfeld2011}%
  \BibitemOpen
  \bibfield  {author} {\bibinfo {author} {\bibfnamefont {W.}~\bibnamefont
  {Rosenfeld}}, \bibinfo {author} {\bibfnamefont {J.}~\bibnamefont {Volz}},
  \bibinfo {author} {\bibfnamefont {M.}~\bibnamefont {Weber}}, \ and\ \bibinfo
  {author} {\bibfnamefont {H.}~\bibnamefont {Weinfurter}},\ }\href {\doibase
  10.1103/PhysRevA.84.022343} {\bibfield  {journal} {\bibinfo  {journal} {Phys.
  Rev. A}\ }\textbf {\bibinfo {volume} {84}},\ \bibinfo {pages} {022343}
  (\bibinfo {year} {2011})}\BibitemShut {NoStop}%
\bibitem [{\citenamefont {Yavuz}\ \emph {et~al.}(2006)\citenamefont {Yavuz},
  \citenamefont {Kulatunga}, \citenamefont {Urban}, \citenamefont {Johnson},
  \citenamefont {Proite}, \citenamefont {Henage}, \citenamefont {Walker},\ and\
  \citenamefont {Saffman}}]{Yavuz2006}%
  \BibitemOpen
  \bibfield  {author} {\bibinfo {author} {\bibfnamefont {D.~D.}\ \bibnamefont
  {Yavuz}}, \bibinfo {author} {\bibfnamefont {P.~B.}\ \bibnamefont
  {Kulatunga}}, \bibinfo {author} {\bibfnamefont {E.}~\bibnamefont {Urban}},
  \bibinfo {author} {\bibfnamefont {T.~A.}\ \bibnamefont {Johnson}}, \bibinfo
  {author} {\bibfnamefont {N.}~\bibnamefont {Proite}}, \bibinfo {author}
  {\bibfnamefont {T.}~\bibnamefont {Henage}}, \bibinfo {author} {\bibfnamefont
  {T.~G.}\ \bibnamefont {Walker}}, \ and\ \bibinfo {author} {\bibfnamefont
  {M.}~\bibnamefont {Saffman}},\ }\href {\doibase
  10.1103/PhysRevLett.96.063001} {\bibfield  {journal} {\bibinfo  {journal}
  {Phys. Rev. Lett.}\ }\textbf {\bibinfo {volume} {96}},\ \bibinfo {pages}
  {063001} (\bibinfo {year} {2006})}\BibitemShut {NoStop}%
\bibitem [{\citenamefont {Jones}\ \emph {et~al.}(2007)\citenamefont {Jones},
  \citenamefont {Beugnon}, \citenamefont {Ga\"etan}, \citenamefont {Zhang},
  \citenamefont {Messin}, \citenamefont {Browaeys},\ and\ \citenamefont
  {Grangier}}]{Jones2007}%
  \BibitemOpen
  \bibfield  {author} {\bibinfo {author} {\bibfnamefont {M.~P.~A.}\
  \bibnamefont {Jones}}, \bibinfo {author} {\bibfnamefont {J.}~\bibnamefont
  {Beugnon}}, \bibinfo {author} {\bibfnamefont {A.}~\bibnamefont {Ga\"etan}},
  \bibinfo {author} {\bibfnamefont {J.}~\bibnamefont {Zhang}}, \bibinfo
  {author} {\bibfnamefont {G.}~\bibnamefont {Messin}}, \bibinfo {author}
  {\bibfnamefont {A.}~\bibnamefont {Browaeys}}, \ and\ \bibinfo {author}
  {\bibfnamefont {P.}~\bibnamefont {Grangier}},\ }\href {\doibase
  10.1103/PhysRevA.75.040301} {\bibfield  {journal} {\bibinfo  {journal} {Phys.
  Rev. A}\ }\textbf {\bibinfo {volume} {75}},\ \bibinfo {pages} {040301}
  (\bibinfo {year} {2007})}\BibitemShut {NoStop}%
\bibitem [{\citenamefont {{Lee}}\ \emph {et~al.}(2018)\citenamefont {{Lee}},
  \citenamefont {{Villa}}, \citenamefont {{Bennett}}, \citenamefont
  {{Stevenson}}, \citenamefont {{Ellis}}, \citenamefont {{Farrer}},
  \citenamefont {{Ritchie}},\ and\ \citenamefont {{Shields}}}]{Lee2018}%
  \BibitemOpen
  \bibfield  {author} {\bibinfo {author} {\bibfnamefont {J.~P.}\ \bibnamefont
  {{Lee}}}, \bibinfo {author} {\bibfnamefont {B.}~\bibnamefont {{Villa}}},
  \bibinfo {author} {\bibfnamefont {A.~J.}\ \bibnamefont {{Bennett}}}, \bibinfo
  {author} {\bibfnamefont {R.~M.}\ \bibnamefont {{Stevenson}}}, \bibinfo
  {author} {\bibfnamefont {D.~J.~P.}\ \bibnamefont {{Ellis}}}, \bibinfo
  {author} {\bibfnamefont {I.}~\bibnamefont {{Farrer}}}, \bibinfo {author}
  {\bibfnamefont {D.~A.}\ \bibnamefont {{Ritchie}}}, \ and\ \bibinfo {author}
  {\bibfnamefont {A.~J.}\ \bibnamefont {{Shields}}},\ }\href@noop {} {\bibfield
   {journal} {\bibinfo  {journal} {ArXiv e-prints}\ } (\bibinfo {year}
  {2018})},\ \Eprint {http://arxiv.org/abs/1804.11311} {arXiv:1804.11311
  [quant-ph]} \BibitemShut {NoStop}%
\bibitem [{\citenamefont {Lindner}\ and\ \citenamefont
  {Rudolph}(2009)}]{Lindner2009}%
  \BibitemOpen
  \bibfield  {author} {\bibinfo {author} {\bibfnamefont {N.~H.}\ \bibnamefont
  {Lindner}}\ and\ \bibinfo {author} {\bibfnamefont {T.}~\bibnamefont
  {Rudolph}},\ }\href {\doibase 10.1103/PhysRevLett.103.113602} {\bibfield
  {journal} {\bibinfo  {journal} {Phys. Rev. Lett.}\ }\textbf {\bibinfo
  {volume} {103}},\ \bibinfo {pages} {113602} (\bibinfo {year}
  {2009})}\BibitemShut {NoStop}%
\bibitem [{\citenamefont {Schwartz}\ \emph {et~al.}(2016)\citenamefont
  {Schwartz}, \citenamefont {Cogan}, \citenamefont {Schmidgall}, \citenamefont
  {Don}, \citenamefont {Gantz}, \citenamefont {Kenneth}, \citenamefont
  {Lindner},\ and\ \citenamefont {Gershoni}}]{Schwartz2016}%
  \BibitemOpen
  \bibfield  {author} {\bibinfo {author} {\bibfnamefont {I.}~\bibnamefont
  {Schwartz}}, \bibinfo {author} {\bibfnamefont {D.}~\bibnamefont {Cogan}},
  \bibinfo {author} {\bibfnamefont {E.~R.}\ \bibnamefont {Schmidgall}},
  \bibinfo {author} {\bibfnamefont {Y.}~\bibnamefont {Don}}, \bibinfo {author}
  {\bibfnamefont {L.}~\bibnamefont {Gantz}}, \bibinfo {author} {\bibfnamefont
  {O.}~\bibnamefont {Kenneth}}, \bibinfo {author} {\bibfnamefont {N.~H.}\
  \bibnamefont {Lindner}}, \ and\ \bibinfo {author} {\bibfnamefont
  {D.}~\bibnamefont {Gershoni}},\ }\href {\doibase 10.1126/science.aah4758}
  {\bibfield  {journal} {\bibinfo  {journal} {Science}\ }\textbf {\bibinfo
  {volume} {354}},\ \bibinfo {pages} {434} (\bibinfo {year}
  {2016})}\BibitemShut {NoStop}%
\end{thebibliography}
%

\end{document}